\documentclass[doublecol]{epl2}

\title{Correlated random fields in dielectric and spin glasses}

\author{M. Schechter\inst{1,2} \and P. C. E. Stamp\inst{1,3}}

\institute{
  \inst{1} Department of Physics and Astronomy, University of
British Columbia - Vancouver, B.C., Canada V6T 1Z1\\
    \inst{2} Department of Physics, Ben Gurion University - Beer Sheva 84105, Israel\\
  \inst{3} Pacific Institute for Theoretical Physics, University of British
Columbia - Vancouver B.C., Canada V6T 1Z1
}

\abstract{Both orientational glasses and dipolar glasses possess an intrinsic
random field, coming from the volume difference between impurity and
host ions. We show this suppresses the glass transition, causing
instead a crossover to the low $T$ phase. Moreover the random field
is correlated with the inter-impurity interactions, and has a broad
distribution. This leads to a peculiar variant of the Imry-Ma
mechanism, with 'domains' of impurities oriented by a few frozen
pairs. These domains are small: predictions of domain size
are given for specific systems, and their possible experimental verification is
outlined. In magnetic glasses in
zero field the glass transition survives, because the random fields
are disallowed by time-reversal symmetry; applying a magnetic field
then generates random fields, and suppresses the spin glass
transition.}

\begin{document}

\maketitle

\newcommand{\bk}{{\bf k}}

\newcommand{\mub}{{\mu_{\rm B}}}
\newcommand{\sD}{{\scriptscriptstyle D}}
\newcommand{\sF}{{\scriptscriptstyle F}}
\newcommand{\sCF}{{\scriptscriptstyle \mathrm{CF}}}
\newcommand{\sH}{{\scriptscriptstyle H}}
\newcommand{\sAL}{{\scriptscriptstyle \mathrm{AL}}}
\newcommand{\sMT}{{\scriptscriptstyle \mathrm{MT}}}
\newcommand{\sT}{{\scriptscriptstyle T}}
\newcommand{\up}{{\mid \uparrow \rangle}}
\newcommand{\down}{{\mid \downarrow \rangle}}
\newcommand{\upt}{{ \langle \uparrow \mid}}
\newcommand{\downt}{{\langle \downarrow \mid}}
\newcommand{\Up}{{\mid \Uparrow \rangle}}
\newcommand{\Down}{{\mid \Downarrow \rangle}}
\newcommand{\bbar}{{\mid \uparrow, 7/2 \rangle}}
\newcommand{\abar}{{\mid \downarrow, 7/2 \rangle}}
\renewcommand{\a}{{\mid \uparrow, -7/2 \rangle}}
\renewcommand{\b}{{\mid \downarrow, -7/2 \rangle}}
\newcommand{\plus}{{\mid + \rangle}}
\newcommand{\minus}{{\mid - \rangle}}
\newcommand{\ex}{{\mid \Gamma_2^l \rangle}}
\newcommand{\LH}{{{\rm LiHoF_4}}}
\newcommand{\LHx}{{{\rm LiHo_xY_{1-x}F_4}}}
\newcommand{\Ht}{{H_{\perp}}}
\newcommand{\Htst}{{H_{\perp}^*}}
\newcommand{\OH}{{\rm OH}^-}
\newcommand{\sL}{{\cal{L}}}
\newcommand{\bfe}{{\bf{e}}}
\newcommand{\bfn}{{\bf{n}}}
\newcommand{\bfR}{{\bf{R}}}

\voffset=-0.18in

Most real solids are either amorphous or disordered crystals, and
many of these show some sort of 'glassy' behavior at low
temperatures\cite{ZP71}. These glasses have been categorized into
different types, including spin\cite{BY86}, structural\cite{PLT02},
electron\cite{DLR82}, orientational\cite{HKL90}, and dipolar
glasses\cite{VG90}. All of these show aging, rejuvenation and memory
effects. However, thermodynamically there is a marked difference
between systems like spin glasses (SGs), which show a clear phase
transition between the paramagnetic and SG phases (with diverging
length scale and consequent diverging nonlinear
susceptibility\cite{BMI81}), and systems like orientational glasses
(OGs) or electric dipole glasses (DG) where the transition to the
low-$T$ phase is smeared\cite{MPA83,SG86,VG90}.

At first glance, this difference seems paradoxical - in OGs, the
rotational degree of freedom behaves like an Ising pseudospin $1/2$,
similar to that in DGs; and the electric- and strain-mediated
interactions between these pseudospins have a similar form to those
between the real spins in dipolar SGs. The quenched randomness then
implies that OGs and DGs should show the same critical behavior as
dipolar SGs, that of the short range Edwards-Anderson
model\cite{BMY86}.

In the present paper we argue that the full answer to this question
is found in an analysis of the effective Hamiltonians for these
systems - including the form of the random fields (RFs) that are
generated - and in the constraints imposed by symmetry. These RFs
are broadly distributed, and correlated with the interactions; we
find this leads to a peculiar disordering mechanism, intermediate
between simple alignment along the RFs and Imry-Ma\cite{IM75}
behavior.

\vspace{3mm}

{\bf (i) Orientational \& Dipolar glasses}: For typical OGs like
KBr:CN, various approaches have been used to describe the
interactions\cite{MR85,Mic86,GRS90a,SS08b}. If one adds to the
standard strain-mediated interaction\cite{GRS90a,BNOK98}
another term reflecting the
volume difference between host and impurity ions, an effective
random field (RF) is found to be generated by the orientational
impurities themselves. The OG degrees of freedom
can be treated formally in the same way as the pseudospin degrees of
freedom in DG systems like Li:KCl, allowing one to analyse the
interactions of both systems on the same footing \cite{SS08b}.

One then finds\cite{SS08b} that over a wide range of temperatures
extending up to the putative glass transition temperature $T_G$,
both OGs and DGs are described by the pseudospin effective
Hamiltonian
\begin{equation}
H_{\rm eff} = \sum_{ij} U_{ij}^{zz}  \hat{\sigma}_i^z \hat{\sigma}_j^z +
\sum_j b_j  \hat{\sigma}_j^z . \label{HeffIsing}
\end{equation}
We emphasize that both the Ising interaction $U_{ij}^{zz}$ and the
RFs $\{ b_j \}$ are generated by the same defect-phonon interaction
\begin{equation}
V_{\rm dp} \;=\; -\sum_j \sum_{\alpha\beta} (\eta_o
\delta^{\alpha\beta} + \gamma_j^{\alpha \beta} \sigma_j^z)
\frac{\partial X_{j \alpha}}{\partial x_{j \beta}}
 \label{V-ph}
\end{equation}
expanded to second order in perturbation. Here $X_{j \alpha}$ represents 
the displacement operator at point $j$ in direction $\alpha$. One 
finds\cite{SS08b}
$U_{ij}^{zz} = g_{zz}{\cal T}^{zz}_{ij} /R_{ij}^3$, and $b_j \equiv
\sum_i U_{ij}^{0z}  = g_{0z} \sum_i {\cal T}^{0z}_{ij}/R_{ij}^3$
with ${\bf R}_{ij} = {\bf r}_i - {\bf r}_j$, and where $g_{zz} \sim
\gamma_o^2/\rho c_o^2$, and $g_{0z} \sim (\eta_o \gamma_o/\rho
c_l^2) $. Here $\rho$ is the density, $c_o (c_l)$ is an average
(longitudinal) sound velocity, and ${\cal T}_{ij}^{zz}, {\cal
T}^{0z}_{ij} \sim O(1)$ are complicated angular averages over the
tensor components of the interaction.

The typical parameters of the Hamiltonian (\ref{HeffIsing}) are
therefore given by
\begin{equation}
\bar{U} \approx \frac{\gamma_o^2}{\rho c_o^2} n \;\; ; \;\; \bar{b}
\approx\; \frac{\eta_o \gamma_o}{\rho c_o^2} n
 \label{bmean}
\end{equation}
and we note that the ratio $\bar{U}/\bar{b}$ is independent of $n$.

In principle, one should add to the hamiltonian (\ref{HeffIsing}) a
tunneling term $\sum_j \Delta_j \hat{\sigma}_j^x$, which is crucial
for the dynamics of the system. However, we are interested in the
case where interactions dominate, ie., $\bar{U}
> \bar{\Delta}$; the tunneling term is then irrelevant when
considering the phase diagram. For concentration $n = 10^{-3}$,
typical values are $\bar{U} \sim 1.5$ K and $\bar{b}$ somewhat
smaller. Typically we assume $n \ll 1$ in what follows, but we will
also compare with larger $n$.

\vspace{3mm}

{\bf (ii) Correlated random fields}:
Consider now the RFs $\{b_j \}$. These were previously argued to
lead to a glassy state\cite{Mic86}, or to a state intermediate
between a true spin glass and a pure random-field
system\cite{RKR04}. Here we argue that actually the RFs destroy
long-range glassy order in the low-T phase\cite{FH86}, which is then
reached by a crossover and not a phase transition. Furthermore,
since the RFs originate from the impurity lattice interactions
themselves, they are correlated with the interactions (because
$U_{ij}^{0z}$ is correlated with $U_{ij}^{zz}$) and
have a broad distribution.
When defects/impurities occupy nearest-neighbour sites, the
field of one on the other is $b_0 \approx \eta_o \gamma_o/(\rho c_o^2) \gg
\bar{b}$.

Because $U_{ij} \propto 1/R^3$ and is random, $\langle U_{ij}^2
\rangle \propto 1/R^6$ and is effectively short
range\cite{BMY86,FH86}. The situation is thus similar to that
analysed by Imry-Ma\cite{IM75}, and its generalisation to glasses by
Fisher and Huse\cite{FH86}. However there is an important difference
- we find that in systems like dilute OGs, where the RFs are broadly
distributed and are correlated with the interactions, one gets
qualitatively different results from systems where the RFs are
normally distributed.
The largest RF, occuring in the event
when defects/impurities occupy nearest-neighbour sites, is $b_0$.
Since $n \ll 1$, such events are rare; however the average RF
$b_{\rm av}$ (the standard deviation of the distribution) is
dominated by these rare events, and is given by $b_{\rm av} \approx
\eta_o \gamma_o \sqrt{n}/(\rho c_o^2) \gg \bar{b}$ (see Ref.~\cite{Sch08},
noting the trivial relations between high and low impurity
densities). Now, since $b_0 \ll U_0$ (where $U_0$ is the interaction
between nearest-neighbour defects/impurities), two regimes are
possible, viz.: (a) $b_0 \ll U_0 n$ and (b) $U_0 n \ll b_0 \ll U_0$.

Regime (a) is simple. All RFs are smaller than the typical
interaction. One can then follow the standard application of the
Imry-Ma argument\cite{IM75} to SGs\cite{FH86}. The energy cost to
flip a domain is related to the typical interaction, and therefore
is $\sim \bar{U} L^{\theta}$, where $L$ is the domain size, and the
stiffness exponent $\theta \approx 0.2$ in 3D\cite{BM84,FH86}.
However, the effective RF is given by  $b_{\rm av} = b_0 \sqrt{n} =
\sqrt{b_0 \bar{b}}$. Thus, the number of spins in a domain $N_d$,
and the related dimensionless correlation length $\tilde{\xi} \equiv
\xi n^{1/3} \sim N_d^{1/3}$ are both $n$ dependent, the latter being
given by
\begin{equation}
\tilde{\xi} \approx \left(\frac{U_0 n}{b_0 \sqrt{n}}\right)^\frac{1}{(3/2) - \theta}
\;\; \approx \;\; n^{0.39} \left(\frac{\bar{U}}{\bar{b}}
\right)^{0.78} .
 \label{corl2}
\end{equation}
In comparison to Gaussian-distributed RFs, the domain size here is
smaller by a factor $n^{0.39}$.

Regime (b) (which usually applies in dilute dipolar glasses) is more
complex, because the spread of RFs is larger than the typical
interactions, so Imry-Ma arguments cannot be used directly. Consider
a pair of nearest-neighbour impurities/defects $l, l'$ (3-impurity
correlations can be neglected when $n \ll 1$). Since $U_0 \gg
\bar{U}$, the two impurities will mutually order, ie., lock
together. If this locking is antiferromagnetic, then the net RF on
the pair comes only from distant impurities - the internally
generated RF $U_{ll'}^{0z} - U_{l'l}^{0z} = 0$, since $U_{ij}^{0z} =
U_{ji}^{0z}$ for any pair $i,j$. However, for ferromagnetic ordering
the internally generated RF is $\sim b_0$, and since $U_0 n \ll
b_0$, this pair will be oriented along this internal RF, and hardly
influenced by the
interactions with other impurities, which have strength $\sim
\bar{U} = U_0 n$! In fact, this is the case for all
ferromagnetically aligned impurity pairs satisfying $b_0/R_{ij}^3 >
\bar{U}$. In a sample of $N$ impurities there are $\approx N
b_0/U_0$ such 'frozen pairs'.

Consider now a sample containing $N$ impurities/defects. The typical
RF resulting from the frozen pairs is then $\sim \sqrt{b_0 N/U_0}
U_0 n$, while that from all other impurities is only $\sim \sqrt{N}
b_0 n$. Thus despite their scarcity, the RF from the frozen pairs
dominates the total RF. Using now the standard argument which
compares the energy gain from the RF and the interaction energy cost
at domain boundaries, we find for a dilute DG in regime (b) a
dimensionless correlation length given by
\begin{equation}
\tilde{\xi} \approx \left(\frac{\bar{U}}{\sqrt{\bar{U}
\bar{b}}}\right)^\frac{1}{(3/2) - \theta} \;\; \approx \;\;
\left({\bar{U} \over \bar{b}} \right)^{0.39} .
 \label{corl1}
\end{equation}

Thus, in this regime the correlation length is determined neither by
the typical RF $b_0 n$, nor by the average RF $b_0 \sqrt{n}$, but by
the algebraic mean of $\bar{U}$ and $\bar{b}$. This is because the
relaxation of the RFs occurs in two stages. The rare pairs with $b_i
> \bar{U}$ relax locally to their frozen state. The majority of the spins
then order via the Imry-Ma mechanism, albeit with RFs dominated by
their interaction with the rare frozen pairs. Because
$\bar{U}/\bar{b}$ is scale-invariant $N_d \sim
(\bar{U}/\bar{b})^{1.17}$, independent of $n$. The low power in the
exponent is dictated by the unique mechanism above, and results in
"quasi-domains", which in practise contain very few
defects/impurities (see the examples below). Moreover, in each such
quasi-domain there are only $\sim (\bar{U}/\bar{b})^{0.17} \approx
1$ frozen pairs. In this limit one cannot of course assume that the
random fields behave as in the large $N$ limit; however, the
mechanism described above gives an upper limit to the size of the
domains, since it minimizes the effect of the RFs. A lower limit can
be given by the assumption that the total RF from the frozen pairs
is maximized, leading to $N_d \sim (\bar{U}/\bar{b})$. In practise
these two limits are indistinguishable, suggesting that domains are
formed in such a way that interactions with frozen pairs are
maximized.

Thus, in regime (b) we arrive at a picture of very small quasi-domains or
'clusters' of $N_d$ pseudospins, of size $\tilde{\xi}$, each
containing $\sim O(1)$ frozen pseudospin, whose field then orients
the rest of them. For $n \approx 1$ (regime (a)) we find $N_d \approx (U/b)^{2.34}$, giving
large clusters, but on strong dilution this
crosses over to the much smaller $N_d \sim (U/b)^{1.17}$. We note
that for all dilutions, the finite $\tilde{\xi}$ gives a finite
nonlinear susceptibility $\chi_3 \propto \tilde{\xi}^{2-\eta}$,
marking a crossover rather than a transition to the glassy
state\cite{GBH94,RY94}, in agreement with experiments\cite{VG90}. However,
the decrease in $N_d$ with $n$ means that the peak in the nonlinear susceptibility
should be much smaller for dilute OGs and DGs, in comparison with
similar systems at high concentrations. This is a central prediction
of this Letter.

In real systems, the ratio $\bar{U}/\bar{b}$ varies widely between
different OGs and DGs, and it is useful to look at some experimental
examples. Consider first $\OH$ impurities in KCl, where the electric
and elastic impurity-impurity interactions are comparable, and
$\bar{U}/\bar{b} \approx 3$ (see Ref.\cite{Dic81}).
In Ref.\cite{SG86} it was argued, based on the behavior of the
non-linear dielectric permittivity of dilute KCl:OH, that there is
no transition to a glass phase, but rather to a state analogous to a
superparamagnet, with roughly $10$ impurities per domain. The
scaling approach we use here, embodied in Eq.~(\ref{corl1}), is
strictly applicable only for $\tilde{\xi} \gg 1$, and even then only
gives an order of magnitude. Thus our result here is consistent with
the above experimental picture of small 'superparamagnetic' domains
for this system. Consider now systems where the spherical and dipolar species are of
similar volume, so $\bar{b}/\bar{U} \ll 1$. For example,
$\bar{b}/\bar{U} \approx 1/20$ in ${\rm KBr:CN}$ (see
Ref.~\cite{YKMP86}), which is an OG for $n \leq 0.5$. For $n \approx
0.5$ we obtain $\tilde{\xi} \approx 10$ and $N_d \approx
(U/b)^{2.34} \approx 1000$. However, in the dilute regime (b),
because of the small exponent in Eq.~(\ref{corl1}), $\tilde{\xi}
\approx 3$, and $N_d \approx 30$ only, a large reduction!

\begin{figure}
\begin{center}
\includegraphics[width = \columnwidth]{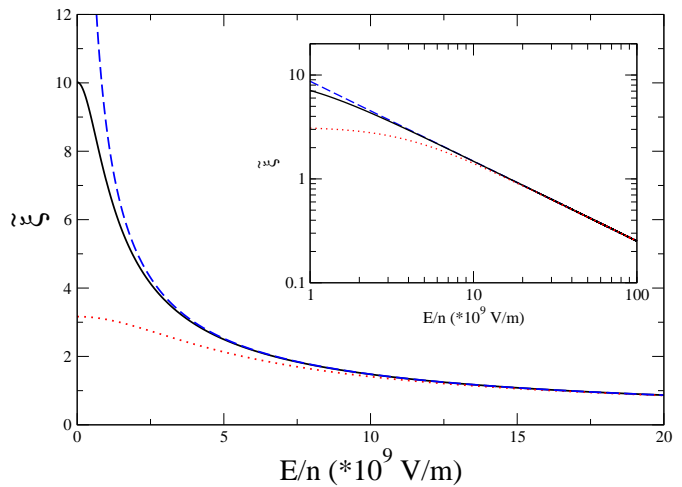} \caption{ The dimensionless correlation length
$\tilde{\xi} \equiv \xi \cdot n^{1/3}$ of the glass order is plotted
as function of the normalized external field $E/n$ for KBr:CN
($\bar{b}/\bar{U} =1/20$, $d = 0.3$ D) with $n=0.5$ (solid black
line) and $n=10^{-3}$ (dotted red line), against its value with zero
effective RF (dashed blue line). The inset shows a log-log plot,
emphasizing the $n$ dependence of the deviation from power low
behavior.}
    \label{CNn}
\end{center}
\end{figure}

As mentioned above, the electric dipole interaction is typically
quite small. For our purposes here, it renormalizes the Ising
interaction, so that
\begin{equation}
\bar{U}_r
\approx n \sqrt{ \left( \frac{\gamma_o^2}{\rho c_o^2} \right)^2 + U_e^2 } \;
 \label{Umean}
\end{equation}
and allows the coupling to an electric field $E$. The latter allows
a measurement of the effective RF as function of $n$, providing an
experimental check to our theory. For $E>0$
\begin{equation}
\tilde{\xi}_E \approx  \left(\frac{\bar{U}}{\sqrt{b_{\rm eff}^2 + (E
\cdot d)^2}}\right)^\frac{1}{(3/2) - \theta} \, .
 \label{Ecorl}
\end{equation}
where $d$ is the dipole moment, $b_{\rm eff}= b_0 \sqrt{n}$ in
regime (a) and $b_{\rm eff}= \sqrt{\bar{b} \bar{U}}$ in regime (b).
Thus, increasing $E$ causes a crossover from impurity-dominated to
field-dominated $\xi$ at $E \cdot d \approx b_{\rm eff}$. In
Fig.~\ref{CNn} we plot $\tilde{\xi}$ for ${\rm KBr:CN}$, for $n =
0.5$ and for $n = 10^{-3} \ll \bar{b}/\bar{U}$.  For $E \gg b_{\rm
eff}/d$, one has $\tilde{\xi}_E \propto E^{-1/(3/2 - \theta)}$.
However, the magnitude of $\tilde{\xi}_E$ at $E=0$, as well as the
region where its functional form deviates considerably from the
above power low, are strongly $n$ dependent. Their measurement as
functions of $n$ would measure the effective RF in the system as
function of dilution, and thus provide a check to our theory.

Extrinsic impurities also generate RFs with a broad distribution,
and can be analyzed along the same lines as the intrinsic
impurities. Thus,
at $E=0$, as the system is purified, the magnitude of $\xi$ and of
the cusp in the nonlinear permittivity increase, but only to a
value in accordance with Eqs. (\ref{corl2}), (\ref{corl1}).

\vspace{3mm}

{\bf (iii) Spin glasses}: We have seen that the random fields
generated in OGs and DGs prevent a genuine phase transition to a
low-$T$ glass state. Why does one then see a glass transition in
SGs? After all, one could certainly expect similar random fields to
be generated therein, via, eg., magnetoacoustic interactions.
However, there is an essential difference between electric and
magnetic Ising systems. If one neglects the volume term in
Eq.~(\ref{V-ph}), then OGs share with SGs a symmetry under $\sigma_z
\rightarrow -\sigma_z$. However this is a symmetry under {\it
parity} $\hat{P}$ in OGs (where the pseudospin variables $\sigma_z$
are not real spins), but under {\it time-reversal} $\hat{T}$ in SGs
(where the $\sigma_z$ are real spins, and couple to real magnetic
fields). As a result, terms linear in $\sigma$ which emerge
naturally in OGs, are not allowed in the absence of a magnetic field
in SGs\cite{noteP}. For example, in zero field the magnetoacoustic
interaction is, to lowest order, bilinear in $\sigma$:
\begin{equation}
V_{sp} \;=\; -\sum_j \sum_{\alpha\beta}\; \left[\eta \delta^{\alpha\beta}
+ \sum_{\gamma \delta} A_j^{\alpha \beta \gamma \delta}
\hat{\sigma}_j^{\gamma} \hat{\sigma}_j^{\delta}\right] \; \frac{\partial
X_{j \alpha}}{\partial x_{j \beta}} ,
 \label{V-SP}
\end{equation}
where $A_j^{\alpha \beta \gamma \delta}$ is the spin-phonon
interaction tensor. Thus, SGs are well described by the Edwards
Anderson model with no RF, and have a well defined phase transition
and diverging nonlinear susceptibility between the SG and PM
phases\cite{BMI81}.

The fact that it is the time-reversal nature of the $Z_2$ symmetry
that prevents the emergence of random longitudinal fields is best
exemplified in anisotropic dipolar magnets. Recently it was shown
that with the application of a transverse field $\Ht$, that breaks
time reversal but by itself keeps the $\sigma_z \rightarrow
-\sigma_z$ symmetry, an effective RF $h$ emerges via the intrinsic
off-diagonal terms of the dipolar interaction\cite{SL06}. Similar RFs would
emerge from the magneto-acoustic interactions in the presence of
$\Ht$. Thus, anisotropic dipolar magnetic
systems in a transverse field are equivalent to dilute electric
systems, albeit with a tunable effective RF. In the SG
phase, these random fields result in a crossover rather than a quantum phase
transition between the SG and PM phases as function of transverse field\cite{SL06}.
Indeed, experiments on the $\LHx$ have found\cite{WBRA93} that the cusp in the nonlinear
susceptibility is smeared, with cusps becoming smaller as temperature is decreased, and
the applied magnetic field at the crossover, and thus the effective random field, are increased.
The above mechanism for RFs
in magnetic system also applies to the FM regime, making
anisotropic dipolar magnets the first
realization of the RFIM in a ferromagnetic
system\cite{Sch08,SL08,SBB+07}. Thus, measuring the RF in anisotropic
 dipolar magnets would be of much interest: this could be done, eg., in the
same way as suggested above for OGs, by mapping $b \leftrightarrow
h$ and $E \leftrightarrow H_{\|}$, where $H_{\|}$ is the
longitudinal magnetic field.

All our considerations for the emergence of effective RFs, both in
OGs and in magnetic systems with transverse magnetic field, are
independent of the thermodynamic phase of the system, i.e. they also
apply to the ferromagnetic/ferroelectric phases\cite{Sch08}. Thus,
we argue that in disordered system with no time-reversal symmetry
the Ising model is unstable to small perturbations, and is therefore
not realizable (i.e., RFs will always emerge). In particular, one
cannot realize the transverse field Ising model in easy-axis
disordered magnetic systems by applying a transverse magnetic field.

In light of these conclusions for SGs, it is interesting to recall
that models of 2-level systems (TLSs) interacting with phonons are commonly
used to describe a wide variety of glasses at low $T$, not just OGs
and DGs. Insofar as the models of TLSs are applicable, and the volume term
in the interaction with phonons causes a breaking of the $\sigma_z
\leftrightarrow -\sigma_z$ symmetry, then our theory predicts that
in these systems RFs will emerge as
well, and there will also be a crossover rather than a phase
transition between the glass-ordered and disordered phases. To
unambiguously test this, one could measure the non-linear dielectric
susceptibility in these systems - it should not diverge as a
function of temperature. These considerations apply both
for the glass transition, and for the low
energy regime of interacting TLSs\cite{note}. For a recent calculation of the strength
of the TLS-TLS interactions in the two regimes in orientational glasses, and a
discussion of its applicability in structural glasses, see ref.\cite{SS09}.

We would like to thank A. Burin, L. Ferrari, AJ Leggett, Z.
Nussinov, B Seradjeh, and AP Young for very useful discussions. This
work was supported by NSERC of Canada and by PITP and CIFAR.


\begin{thebibliography}{10}

\bibitem{ZP71}
Zeller R. C. and Pohl R. O., Phys. Rev. B {\bf 4} (1971) 2029.

\bibitem{BY86}
Binder K. and Young A. P., Rev. Mod. Phys. {\bf 58} (1986) 801.

\bibitem{PLT02}
Pohl R. O., Liu X. and Thompson E., Rev. Mod. Phys {\bf 74} (2002) 991.

\bibitem{DLR82}
Davies J. H., Lee P. A. and Rice T. M., Phys. Rev. Lett. {\bf 49} (1982) 758.

\bibitem{HKL90}
Hochli U. T., Knorr K. and Loidl A., Adv. Phys. {\bf 39} (1990) 405.

\bibitem{VG90}
Vugmeister B. E. and Glinchuk M. D., Rev. Mod. Phys. {\bf 62} (1990) 993.

\bibitem{BMI81}
Barbara B., Malozemoff A. P. and Imry Y., Phys. Rev. Lett. {\bf
47} (1981) 1852; Omari R., Prejean J. J. and Souletie J., J.
Physique {\bf 44} (1983) 1069.

\bibitem{MPA83}
Moy D., Potter R. C. and Anderson A. C., J. Low Temp. Phys. {\bf
52} (1983) 115.

\bibitem{SG86}
Saint-Paul M. and Gilchrist J. le G., J. Phys. C {\bf 19} (1986) 2091.

\bibitem{BMY86}
Bray A. J., Moore M. A. and Young A. P., Phys. Rev. Lett. {\bf 56} (1986) 2641.

\bibitem{IM75}
Imry Y. and Ma S. K., Phys. Rev. Lett. {\bf 35} (1975) 1399.

\bibitem{MR85}
Michel K. H. and Rowe J. M., Phys. Rev. B {\bf 32} (1985) 5818,
{\it ibid.} (1985) 5827.

\bibitem{Mic86}
Michel K. H., Phys. Rev. Lett. {\bf 57} (1986) 2188.

\bibitem{GRS90a}
Grannan E. R., Randeria M. and Sethna J. P., Phys. Rev. B {\bf 41}, (1990) 7784, ibid. {\bf 41} (1990) 7799.

\bibitem{SS08b}
Schechter M. and Stamp P. C. E., J. Phys. Condens. Matter {\bf 20} (2008) 244136.

\bibitem{BNOK98}
Burin A. L., Natelson D., Osheroff D. D. and Kagan Y., in {\it
Tunneling Systems in Amorphous and Crystalline Solids}, edited by
Esquinazi P. (Springer, Berlin) 1998.

\bibitem{RKR04}
Rheinstadter M. C., Knorr K. and Rieger H., Phys. Rev. B {\bf 69} (2004) 144427.

\bibitem{FH86}
Fisher D. S. and Huse D. A., Phys. Rev. Lett. {\bf 56} (1986) 1601;
Phys. Rev. B {\bf 38} (1988) 386.

\bibitem{Sch08}
Schechter M., Phys. Rev. B {\bf 77} (2008) 020401(R).

\bibitem{BM84}
Bray A. J. and Moore M. A., J. Phys. C {\bf 17} (1984) L463;
McMillan W. L., Phys. Rev. B {\bf 30} (1984) R476.

\bibitem{GBH94}
Guo M., Bhatt R. N. and Huse D. A., Phys. Rev. Lett. {\bf 72} (1994) 4137.

\bibitem{RY94}
Rieger H. and Young A. P., Phys. Rev. Lett. {\bf 72} (1994) 4141.

\bibitem{Dic81}
Dick B. G., Phys. Rev. B {\bf 24} (1981) 2127.

\bibitem{YKMP86}
De Yoreo J. J., Knaak W., Meissner M. and Pohl R. O., Phys. Rev. B
{\bf 34} (1986) 8828.

\bibitem{noteP}
We would like to thank Peter Young for a discussion of this point.

\bibitem{SL06}
Schechter M. and Stamp P. C. E., Phys. Rev. Lett. {\bf 95} (2005) 267208; Schechter M. and Laflorencie N., Phys. Rev. Lett. {\bf 97} (2006) 137204; Schechter M., Stamp P. C. E. and Laflorencie N., J.
Phys.: Condens. Matter {\bf 19} (2007) 145218.

\bibitem{WBRA93}
W. Wu, D. Bitko, T. F. Rosenbaum, and G. Aeppli, Phys. Rev. Lett.
{\bf 71}, (1993) 1919.

\bibitem{SBB+07}
Silevitch D. M., Bitko D., Brooke J., Ghosh S., Aeppli G. and Rosenbaum T. F., Nature {\bf 448} (2007) 567.

\bibitem{SL08} Schechter M. and Stamp P. C. E., Phys. Rev. B {\bf 78} (2008) 054438.

\bibitem{note}
A transition to coherent orbital motion involving pairs of TLS coupled to nuclear quadrupolar moments is apparently seen at mK in some dipolar glasses (see Refs. \cite{SEH98,NFHE04}). However this is not a glass transition.

\bibitem{SEH98} Strehlow P., Enss C. and Hunklinger S., Phys. Rev. Lett. {\bf 80} (1998) 5361.

\bibitem{NFHE04} Nagel P., Fleischmann A., Hunklinger S. and Enss C., Phys. Rev. Lett. {\bf 92} (2004) 245511.

\bibitem{SS09} Schechter M. and Stamp P. C. E., arXiv:0910.1283.

\end{thebibliography}
\end{document}